\documentclass[12pt,preprint]{aastex}

\shorttitle{Optical Variability of J~1128+5925}
\shortauthors{Wu et al.}

\received{}
\accepted{}

\begin{document}

\title{Optical Variability of the Radio Source J~1128+5925 -- II. Confirmation
       of Its Optical Quietness}

\author{Jianghua Wu, Xu Zhou, Jun Ma, Zhenyu Wu, Zhaoji Jiang, Jiansheng Chen}
   \affil{National Astronomical Observatories, Chinese Academy of Sciences,
        20A Datun Road, Chaoyang District, Beijing 100012, China}
   \email{jhwu@bao.ac.cn}

\begin{abstract}
The source J~1128+5925 was found recently to show strong intraday variability
at radio wavelengths and its radio variability may come from interstellar
scintillation. In optical, the object was quiet in our 2007 monitoring
session. Here we report the results of our new optical
monitoring of this source in 2008. In addition to confirm our 2007
results, that the object did not display any clear variation on timescales
from hour--day to month, we provide evidence that the object does not vary
on timescale of one year, and it is probably intrinsically quiet in
optical domain. Its very different behaviors in optical and radio regimes can
be naturally explained if its strong radio variability comes from interstellar
scintillation.
\end{abstract}

\keywords{galaxies: active --- quasars: individual (J 1128+5925)}

\section{INTRODUCTION}
Blazars are believed to be those active galactic nuclei (AGNs) with their
relativistic jets pointed basically to our line of sight \citep{antonucci93,
urry95}. Because of the beaming and relativistic effects in the jet, blazars
show strongest and fastest variability among all AGNs, and this variability
show up across the entire electromagnetic spectrum. The monitoring of blazars
simultaneously at multi-wavelength can reveal their spectral energy
distributions, and is crucial to constrain their emission and variation
mechanisms \citep[e.g.,][]{wagner96,ghisellini97,taglia03,boettcher04,
boettcher07,boettcher08}. Based on the strength of their emission lines,
blazars are currently separated into BL Lacertae objects and flat-spectrum
radio quasars \citep[e.g.,][]{marcha96,landt04}.

Recently, the flat-spectrum radio quasar J~1128+5925 was found to show strong
intra-day variability (IDV) at centimeter wavelengths, and its IDV timescale
displayed an annual modulation \citep{gabanyi07}. Therefore, its radio
IDV may come from interstellar scintillation \citep[ISS, for a review
of various variation mechanisms, see][]{wagner95}. In order to open a new
window to investigate the variability of this object, we made the first
optical monitoring of J~1128+5925 in 2007 May. We found that the object only
showed trivial variability on internight timescale and did not present any
clear intranight variability \citep[][hereafter Paper I]{wu08}.

However, our 2007 monitoring session was not long enough to determine whether
the object was just quiet at that period of time or is intrinsically quiet in
optical regime. In order to resolve this ambiguity, we carried out a new
monitoring program on this object in 2008. Here we present our observations, 
data reduction procedure, and the results in the next sections. Discussions 
on the variation mechanisms are given at the end.

\section{OBSERVATIONS AND DATA REDUCTIONS}
The monitoring was performed with a 60/90 cm f/3 Schmidt telescope at Xinglong
Station, National Astronomical Observatories of China. The telescope
is equipped with a $4096\times4096$ E2V CCD, which has a pixel size of
$12\,\micron$ and a spatial resolution of $1.3\,\arcsec\rm{pixel}^{-1}$. When
used for blazar monitoring, only the central $512\times512$
pixels are read out as a frame. Each such frame has a field of view of about
$11'\times11'$. Our monitoring was made in the Cousins $R$-band, and covered
the period from 2008 April 22 to June 4. The exposure times are mostly 480 s
and can be as short as 120 s, depending on weather and moon phase.

\begin{figure}
\plotone{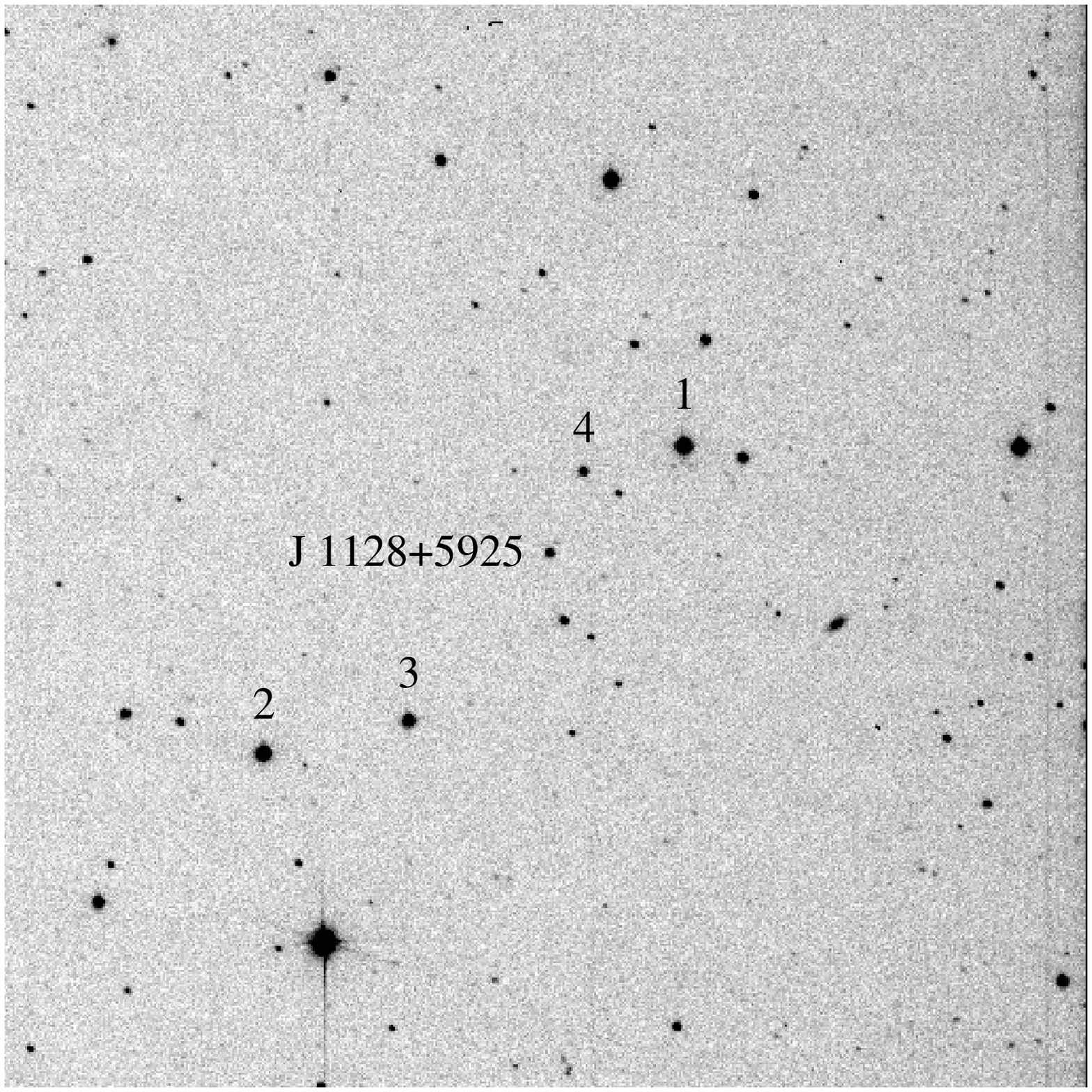}
\caption{Finding chart of J~1128+5925. The blazar and four comparison stars
are labeled. Upwards is north, left is east.}
\label{F1}
\end{figure}

The data reduction procedures include bias subtraction, flat-fielding,
extraction of instrumental aperture magnitude, and flux calibration. We used
differential photometry. For each frame, the instrumental magnitudes of the
blazar and four comparison stars (See Fig.~\ref{F1}) were extracted at first.
The radii of the aperture and the sky annuli were adopted as 3, 7, and 10
pixels, respectively. Then the brightness of the blazar was measured relative
to the average brightness of the three reference stars 1, 2, and 3. Star 4
acted as a check star, which has an apparently similar brightness as the
blazar \citep[for a reasonable selection of reference and check stars, see][]
{howell88}. Its brightness was also measured relative to the average
brightness of the three reference stars, so as to verify the stable fluxes of
the four comparison stars, and to verify the accuracy of our measurements.
This definition of reference and check stars is different from that
in Paper I. In Paper I, the check star was among the reference stars. When
the brightness of check star was measured relative to the average of all
reference stars, it was actually measured relative partly to itself, and
would result in an under-estimated `variation' of the check star. This
is unreasonable and was rejected in this paper.\footnote{The conclusion that
J~1128+5925 didn't vary on all nights in Fig.~3 of Paper I is still viable
because the quietness of this object was justified by even a higher-than-actual
standard ($\sigma_{\rm B}$ was slightly under-estimated) on these nights.}

\section{RESULTS}
\begin{figure}
\plottwo{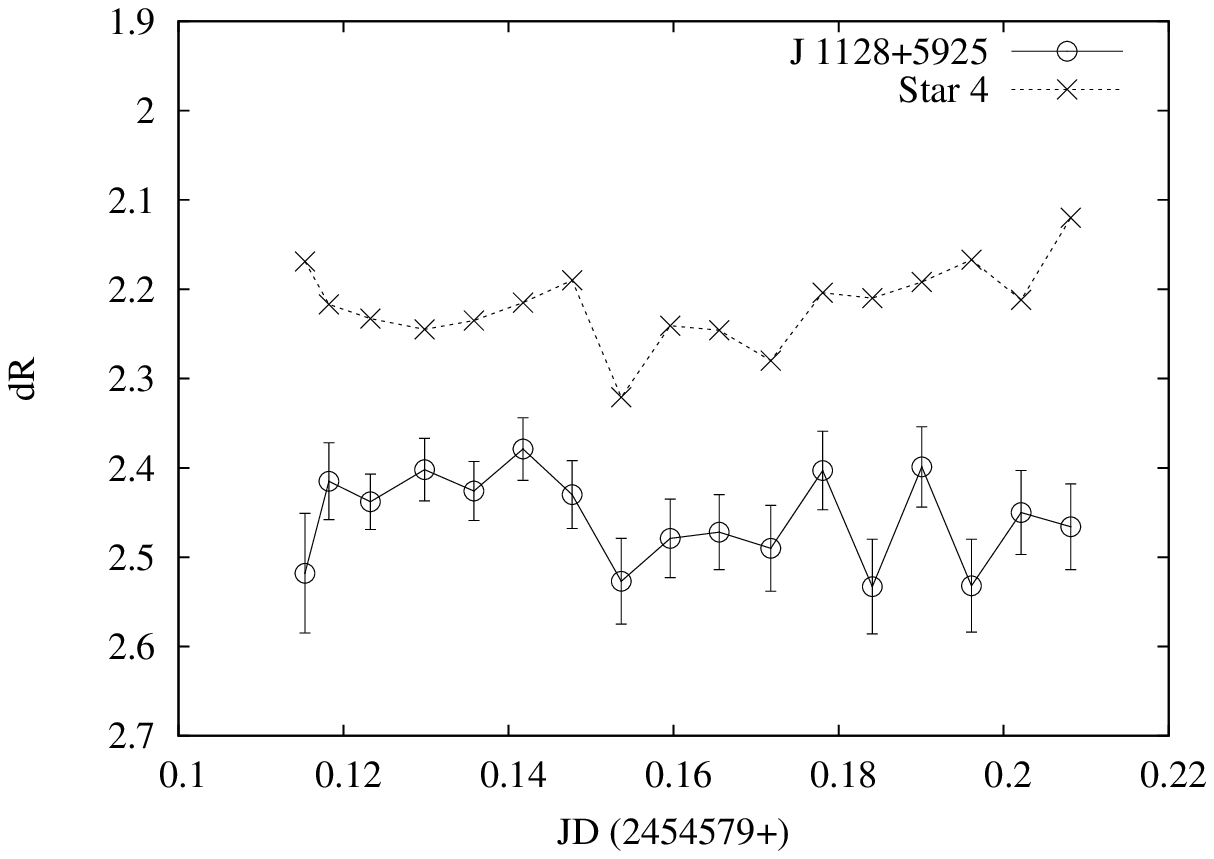}{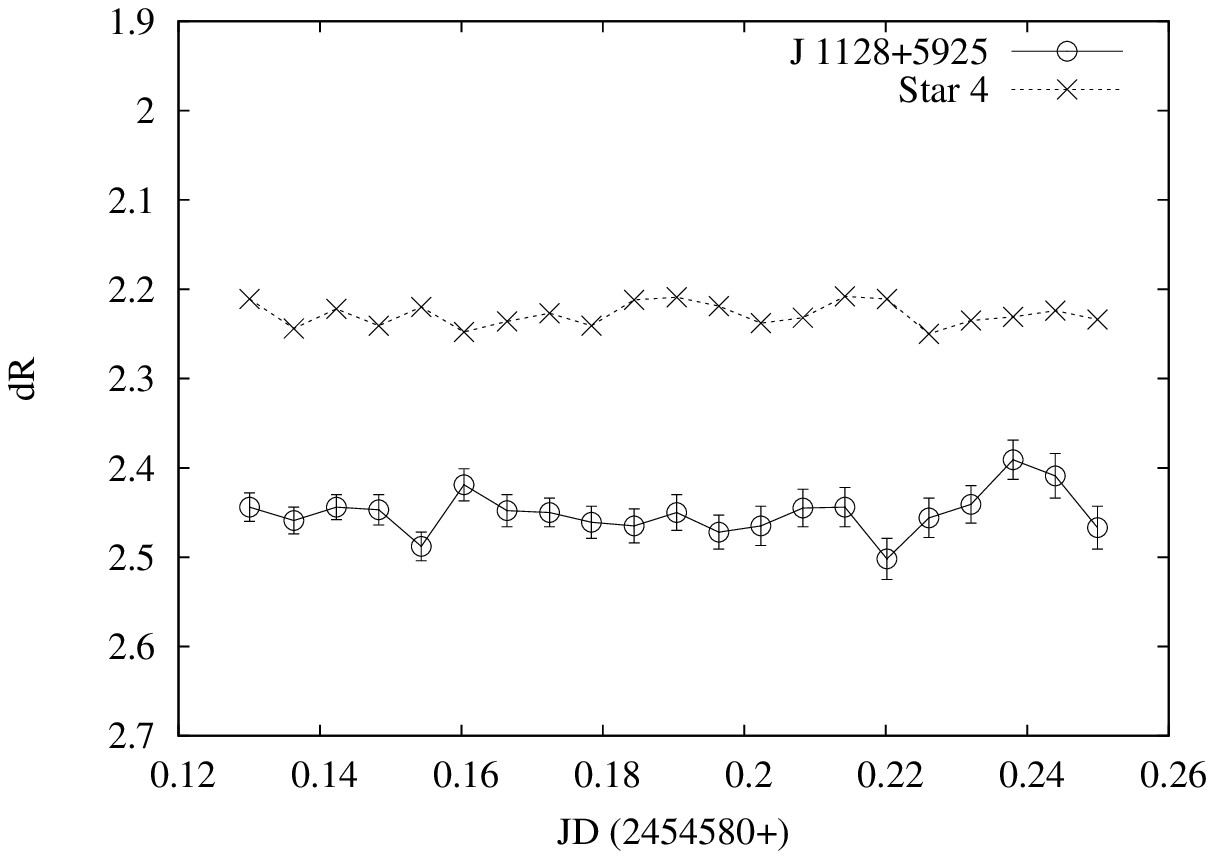}
\plottwo{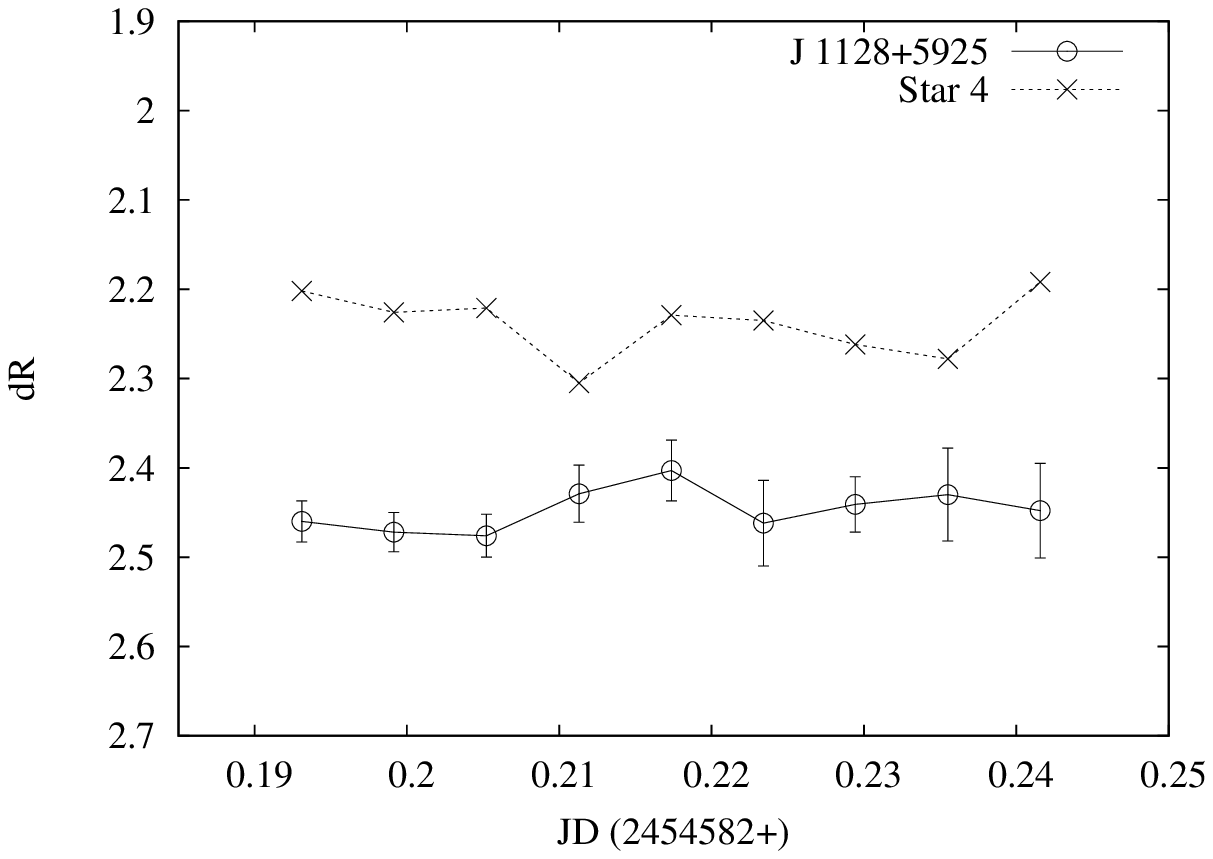}{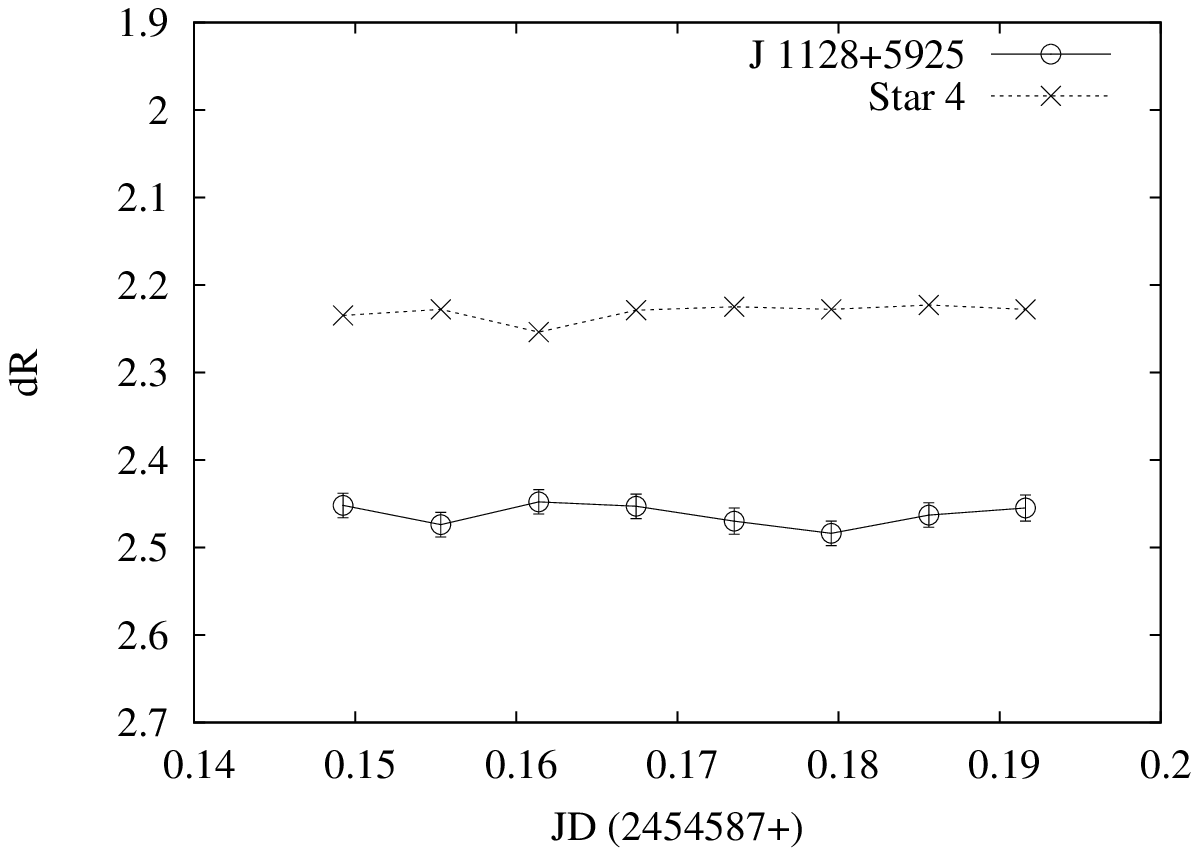}
\plottwo{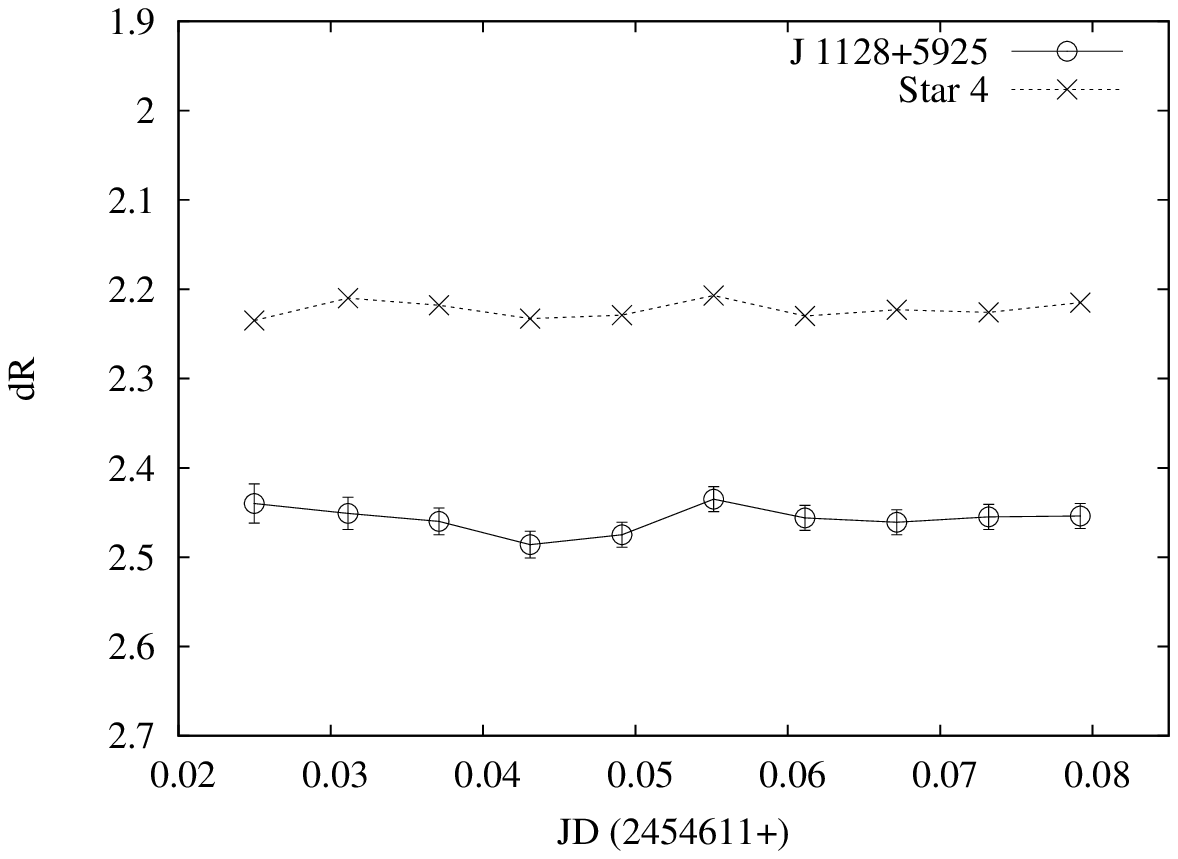}{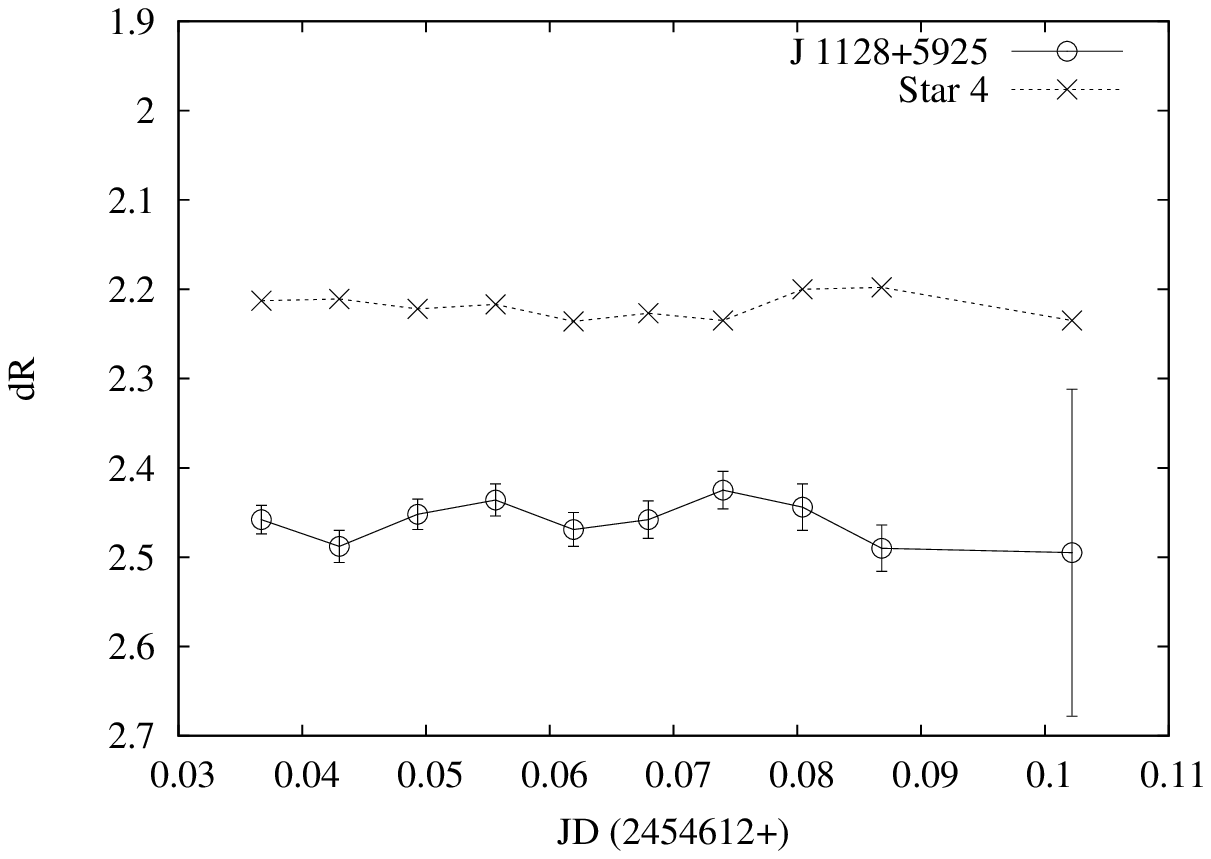}
\epsscale{0.45}
\plotone{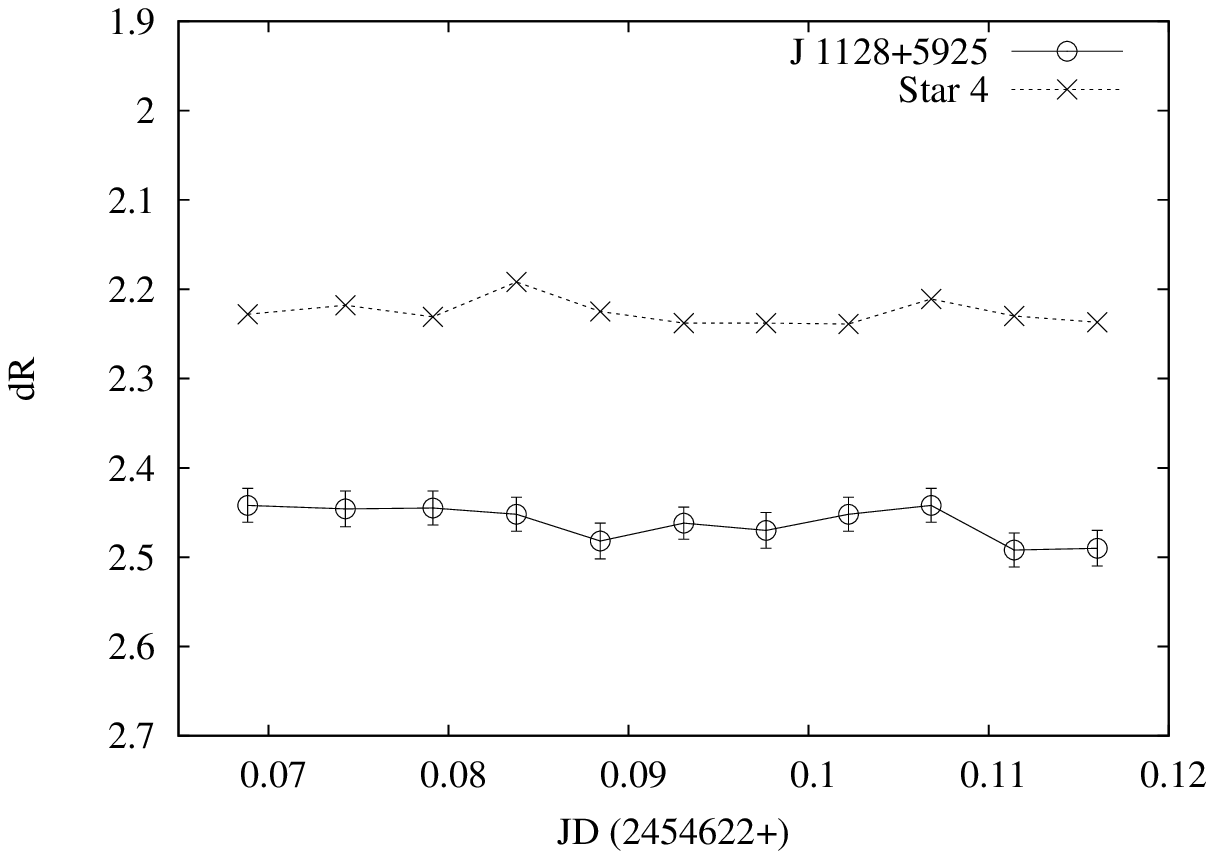}
\caption{Intranight light curves on seven nights with intensive monitorings.
The open circles and solid lines show the light curves of the blazar, while
the crosses and dotted lines show the light curves of star 4.}
\label{F2}
\end{figure}

Figure~\ref{F2} shows the intranight light curves of J~1128+5925 on seven
nights with intensive monitorings. Also shown are those of star 4, the check
star. Except for the first panel (JD 2,454,579), where the light curves have
relatively large errors, all other panels have light curves close to
horizontal straight lines, and do not demonstrate any clear variation.

A quantitative assessment was performed on whether or not the object was
variable on these seven nights. As in \citet{jang97}, \citet{stalin06},
\citet{hu06}, and Paper I, a parameter $C$ is defined as
$C=\sigma_{\rm B}/\sigma_{\rm S}$, where $\sigma_{\rm B}$ is the standard
deviation of the magnitudes of the blazar and $\sigma_{\rm S}$ is that of
the check star. When $C\geqslant2.576$, the object can be claimed to be
variable at the 99\% confidence level. Table~1 lists the results. All $C$'s
are less than 2.0, implying that J~1128+5925 was not variable on these
nights.

\begin{figure}
\epsscale{1.0}
\plottwo{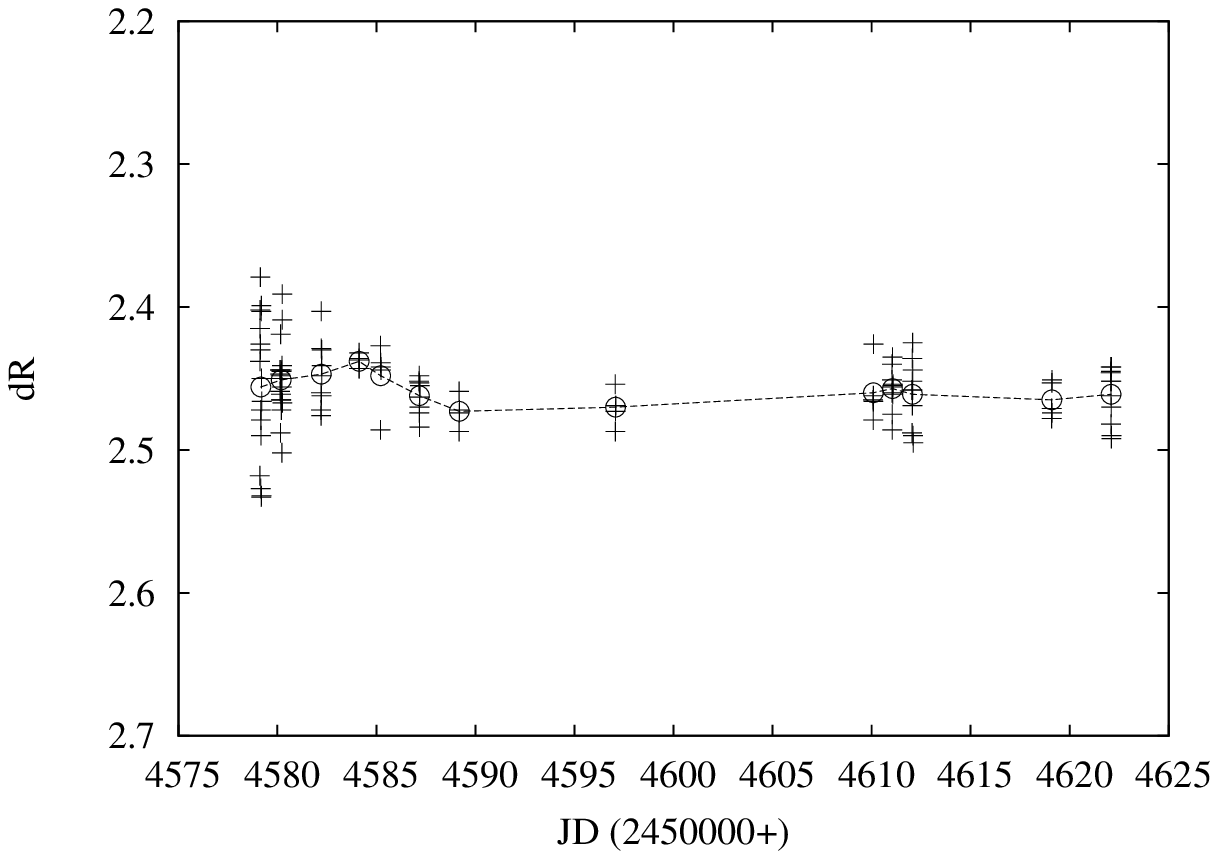}{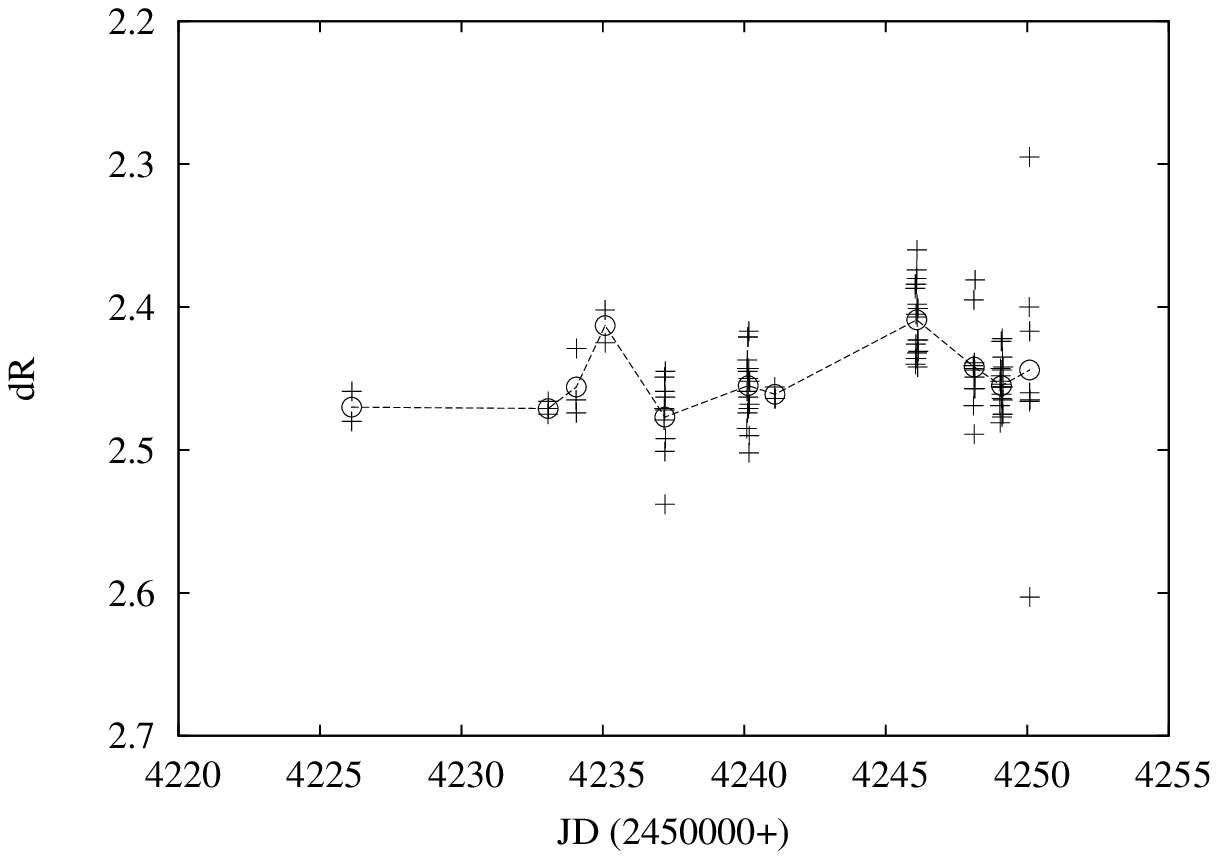}
\caption{The light curves of all nights in 2008 ({\sl left}) and 2007 ({\sl
right}). The plus symbols are individual measurements, while the open circles
and dashed lines mark the nightly average light curves.}
\label{F3}
\end{figure}

The light curve of the whole monitoring period is shown in Figure~\ref{F3}
({\sl left}). Across the period of 44 days, the nightly average brightness of
the blazar, as indicated by the open circles and the dashed line, was quite
stable at $dR\sim 2.46$, and had an amplitude of 0.04 mags and a R.M.S of 0.01
mags.  When we take into account the probability that the peak around JDs
2,454,584 and 2,454,585 and the trough from JDs 2,454,589 to 2,454,597 might
be biased by very few observations on these nights, the amplitude and R.M.S
may be even smaller. This is to say, the object was quiet on a timescale of
about one month.

As a comparison, the light curve of our 2007 monitoring session, which
covered a period of 25 days, is also displayed in Figure~\ref{F3} ({\sl right}).
The magnitudes were re-calibrated by using the new definition of comparison
stars described in \S2. The average brightness is
$dR\sim2.45$, the same as in 2008. This similarity is unlikely by chance,
because our 2007 and 2008 monitoring sessions lasted for close to and more
than a month, respectively. Therefore, the object may not vary on a timescale
of one year.

\section{CONCLUSIONS AND DISCUSSIONS}
We conducted a new optical monitoring program on radio source J~1128+5925
in the Cousins $R$-band from 2008 April 22 to June 4. The results show that
the object still did not demonstrate any clear variability on timescales from
hour--day to month, which confirmed our 2007 monitoring results in Paper I.
Furthermore, the object had similar brightness in 2008 as in 2007, suggesting
that it may not vary even on a timescale of one year. Being quiet on such a
long timescale, the object was unlikely just in a quiescent state at optical
wavelength in our two monitoring sessions. It may be intrinsically quiet in
optical domain.

In paper I, we proposed simultaneous radio and optical observations on this
source in order to verify the extrinsic origin of its radio IDV. Now our new
optical observations reinforced its optical quietness. In radio, there were
respectively about three days of observations on this source in 2008 April
and June, and it still illustrated strong IDV phenomena (T. Krichbaum, private
communication), as in \citet{gabanyi07}. Because of the very limited time
coverage by both optical and radio observations, a correlation analysis was
impractical. In fact, the very different behaviors in optical and radio
regimes ruled out the probability of correlated optical and radio variations
in J~1128+5925.

The origin of the IDV in radio sources is still a subject of much debate. If
the origin is intrinsic, the short timescale would require a very small
emission region and hence an extremely high apparent brightness temperature
of $10^{16}-10^{21}$ K, which is far beyond the inverse-Compton limit of
$\sim10^{12}$ K \citep{keller69}. Then high Doppler factors up to 40--100 are
required to reconcile this contradiction \citep[e.g.,][and references therein]
{peng00}. Extrinsic origins, such as ISS, can also result in IDV in radio
sources \citep{rickett90}. Indeed, strong IDV can be convincingly explained in
terms of ISS in some blazars, such as J~1819+3845 \citep{dennett02,dennett03}
and PKS~1257$-$326 \citep{bignall03}. However, some other blazars, such as
S5~0716+714,
show correlated IDVs in radio and optical regimes \citep{quirren91,wagner96}.
This is taken as strong evidence for the intrinsic origin of IDV, because
the ISS cannot result in variability in optical regime \citep{wagner95}.
J~1128+5925 demonstrates strong IDV at radio wavelengths but is very quiet
in optical domain. This can be naturally explained if the strong IDV comes
from ISS, which was suggested by \citet{gabanyi07}.

\begin{acknowledgements}
This work has been supported by the Chinese National Natural Science
Foundation grants 10603006, 10573020, 10633020, 10873016, 10803007, and by
National Basic Research Program of China (973 Program) No. 2007CB815400.
\end{acknowledgements}

\clearpage

\begin{deluxetable}{ccccc}
\tablewidth{0pt}
\tablecaption{Statistics on Seven Intranight Light Curves}
\tablehead{ \colhead{Julian Date} & \colhead{N} & \colhead{Duration (h)} &
\colhead{$C$} & \colhead{Var?} \\ }
\startdata
2,454,579 & 17 & 2.23 & 1.11 & N \\
2,454,580 & 21 & 2.88 & 1.83 & N \\
2,454,582 &  9 & 1.15 & 0.65 & N \\
2,454,587 &  8 & 1.02 & 1.29 & N \\
2,454,611 & 10 & 1.32 & 1.54 & N \\
2,454,612 & 10 & 1.58 & 1.69 & N \\
2,454,622 & 11 & 1.13 & 1.33 & N \\
\enddata

\end{deluxetable}

\end{document}